\documentclass[a4paper,11pt]{article}
\usepackage{pos}

\title{Sensitivity of the Cherenkov Telescope Array to emission from the gamma-ray counterparts of
neutrino events}
 \ShortTitle{CTA NToO}

\author*[a]{Olga Sergijenko}
\author[b]{Anthony M. Brown} 
\author[c]{Damiano Fiorillo}
\author[b]{Alberto Rosales de Le\'{o}n} 
\author[d]{Konstancja Satalecka}
\forColl{CTA Consortium} 
\affiliation[a]{Astronomical Observatory, Taras Shevchenko National University of Kyiv, Kyiv, Ukraine}
\affiliation[b]{Centre
for Advanced Instrumentation (CfAI), Department of Physics, University of Durham, Durham, UK}
\affiliation[c]{Universita degli Studi di Napoli “Federico II” and INFN - Sezione di Napoli, Napoli, Italy}
\affiliation[d]{DESY, 15738 Zeuthen, Germany}

\author[e]{Chun Fai Tung} 
\author[f]{René Reimann} 
\author[g]{Theo Glauch} 
\author[e]{Ignacio Taboada}
\forColl{ and the FIRESONG Team}
\affiliation[e]{Center for Relativistic Astrophysics and School of Physics, Georgia Institute of Technology,
Atlanta, GA 30332, USA}
\affiliation[f]{Institut of Physics, Johannes Gutenberg University Mainz, 55128
Mainz, Germany}
\affiliation[g]{Technische Universität München, Physik-Department,
D-85748 Garching bei München, Germany}

\emailAdd{olga.sergijenko.astro@gmail.com, anthony.brown@durham.ac.uk, damianofg@gmail.com,
alberto.rosales-de-leon@durham.ac.uk, konstancja.satalecka@desy.de, ctung6@gatech.edu, nerreimann@uni-mainz.de,
theo.glauch@tum.de, itaboada@gatech.edu}

\abstract{We investigate the possibility of detection of the VHE gamma-ray counterparts to the neutrino astrophysical sources within the Neutrino Target of Opportunity (NToO) program of CTA using the populations simulated by the FIRESONG software to resemble the diffuse astrophysical neutrino flux measured by IceCube. We derive the detection probability for different zenith angles and geomagnetic field configurations. The difference in detectability of sources between CTA-North and CTA-South for the average geomagnetic field is not substantial. We investigate the effect of a higher night-sky background and the preliminary CTA Alpha layout on the detection probability.
}

\FullConference{37$^{\rm{th}}$ International Cosmic Ray Conference (ICRC 2021)\\
		July 12th -- 23rd, 2021\\
		Online -- Berlin, Germany}

\begin{document}
\maketitle

\section{Introduction}
The Cherenkov Telescope Array will be the next generation ground-based imaging atmospheric Cherenkov telescope (IACT) array. Building upon the strengths of the current generation of telescopes, such as H.E.S.S., MAGIC and VERITAS, CTA will detect gamma rays from the energy of 20 GeV to above 300 TeV, with an order of magnitude improvement in sensitivity, as well as improved angular and energy resolution compared to the current generation IACTs. To achieve this improved performance CTA will consist of three telescope sizes, small, medium and large (SST, MST and LST respectively), with each one optimised for a specific energy range. 
CTA will consist of two arrays, one in each hemisphere, with the northern array (CTA-N) consisting of 13 telescopes spread over 0.5 km$^2$ in La Palma, Spain, and the southern array (CTA-S) consisting of 51 telescopes spread over 4~km$^2$ near Paranal, Chile. 

In the context of the origin of transient astrophysical messengers, like neutrinos, CTA has several important performance characteristics. Firstly, CTA telescopes are designed to rapidly re-position to any location in the sky. As a point of reference, the LSTs can re-position to anywhere in the sky above $30^{\circ}$ in just 20 seconds, thus minimising the time between receiving alerts from other observatories, such as IceCube, and starting observations. Secondly, the field-of-view (FoV) for LSTs will be $\sim4^{\circ}$, for MSTs $\sim8^{\circ}$ and for SSTs $\sim10^{\circ}$. 

Astrophysical sources capable of hadronic acceleration to relativistic energies have long been believed to be sources of astrophysical neutrinos, with the neutrinos originating from the decay of charged mesons created by the relativistic hadrons interacting with ambient gas and/or ambient radiation within the astrophysical source. The decay of these mesons results in the emission of neutrinos and gamma rays. As such, assuming that the photon opacity of the emission region is low enough to allow gamma rays to escape, searching for spatially and temporally correlated gamma-ray and neutrino emission allows us to constrain the origin of the astrophysical neutrinos observed by the IceCube Neutrino Observatory.

In April 2016, IceCube initiated a \textit{Realtime alert program} in which neutrinos with high probability of being of astrophysical origin are reported in realtime via the Gamma-ray-burst Coordinate Network (GCN) \cite{RealtimeIceCube}. The program reports $\nu_\mu$ candidate events, which have the advantage of well-localized angular positions, typically $\Delta\psi \sim 0.5^\circ$. Starting on June 17, 2019, a new version of the realtime program became operational \cite{Tung2019}. In this updated system, two streams of neutrinos are provided. The \textit{Gold} stream delivers $\sim$10 events per year with an average astrophysical signal purity of 50\%. The \textit{Bronze} stream has an average purity of 30\% for an additional $\sim$20 alerts per year. 

In the Key Science Project proposed in the CTA Science Book \cite{scta} a total of 5 hours per site, per year, has been allocated to high-energy neutrino events observations, during the first 10 years of CTA operation. In these proceedings, we outline the ongoing work to quantify the efficiency with which CTA will observe gamma-ray emission from IceCube `Gold' $\nu_\mu$ alerts and all-sky `hot-spots' (i.e. highest significance positions). The structure of the proceedings is as follows: in Section 2 we outline the simulation tool set we used to simulate IceCube alerts, in Section 3 we outline our conversion from the neutrino flux to the expected gamma-ray flux. Section 4 describes our simulations of CTA response to the expected gamma-ray flux, while in Section 5 we state our results.


\section{FIRESONG}
The sources of astrophysical high-energy neutrinos have not been unequivocally identified. Observations of TXS 0506+056 provide evidence that blazars or a sub-population of blazars are possibly responsible for a fraction of the neutrino flux \cite{IceCube:2018cha, IceCube:2018dnn}. Nevertheless, evidence has only been claimed for one blazar/neutrino correlation therefore one has to be careful while extending its properties to describe the whole population of neutrino sources. The population of potential neutrino sources responsible for this flux can be parametrized in terms of local density (local density rate) vs. neutrino luminosity (flare energy) for steady (flaring) sources~\cite{Kowalski2014}. 

FIRESONG \cite{firesong_ref} is an open-source software that simulates neutrino source populations for a given local density and neutrino luminosity, by making additional assumptions on the source density evolution 
and their luminosity functions. 
In these proceedings we explore a source population which does not evolve with redshift (which is a simplified way to describe blazars) and one that follows the star formation rate (SFR) evolution. In both populations, sources are described as standard candles, i.e., all sources share the same neutrino luminosity. 
We also assume that the specific simulated class is responsible for 100\% of the astrophysical neutrino flux \cite{Aartsen:2017mau}. FIRESONG is able to describe alerts with the same characteristics as those reported by IceCube. It should be noted that the rate of alert and its declination dependence depend on the IceCube response, notably the effective area. However, the redshift distribution of the sources of alerts depends on the cosmology model and the properties of the assumed population. For this work we have used  the $\Lambda$CDM cosmological model with the parameters derived from the Planck-2015 data \cite{planck2015}: $\Omega_{M} =  0.308$, $\Omega_{\Lambda} = 0.692$ and $h = 0.678$.


\subsection{TXS-like flaring sources}
The first simulated scenario is modeled after the neutrino flare of TXS 0506+056 (TXS) in the season 2014-2015. We follow the model suggested by \cite{Halzen2019} in which only a fraction of blazars, including TXS, are responsible for the astrophysical neutrino flux. We will call this population \textit{TXS-like sources}. The local burst density rate can be related to the local density of blazars as:
\begin{equation}
    \dot{\rho} = F \times \rho_{\mathrm{BL}}.
\end{equation}
where $F$ is a fraction of the total number of blazars. As described before, we assume these sources are neutrino standard candles and display no evolution in the density rate. 
We also assume all TXS-like sources have the same flare duration in their local reference frame. Correcting the 110 days flare of TXS~\cite{IceCube:2018cha} for redshift, this corresponds to 82 days. Saturating the astrophysical neutrino flux with these flares results in a uniquely defined burst energy as a function of $F$. 


\subsection{Steady sources}
The second simulated scenario is that the diffuse neutrino flux is due to steady neutrino sources. Although IceCube has not yet resolved a point source above 5$\sigma$ significance, there are sources that have exceeded the IceCube sensitivity (Note that sensitivity in IceCube has a different definition than for IACTs; see Ref. \cite{Aartsen:2018ywr}). These sources will be of interest for the NToO program of CTA. Assuming the neutrino sources are i) standard candles, ii) follow the star formation evolution from \cite{Madau:2014bja} or a flat redshift evolution, and iii) saturate the astrophysical neutrino flux \cite{Aartsen:2017mau}, we can simulate the whole-sky source population with different local density conditions. 
Then, the flux from each neutrino source is compared with IceCube sensitivity \cite{Aartsen:2018ywr}. The sources that exceed IceCube sensitivity are used as seeds for the NToO for CTA. 


\section{VHE gamma-ray emission accompanying the neutrino emission}
In order to calculate the gamma-ray flux emitted together with neutrinos we assume that they are produced in proton interactions with the surrounding photon field ($p\gamma$ interactions) as usually postulated for Active Galactic Nuclei (AGN), and do not consider any additional absorption or cascading of $\gamma$ rays inside the source. The secondary pions and other particles decay to neutrinos or gamma rays and in the simplest case the relation between the gamma ray and neutrino production rates is: 
\begin{equation}
    \frac{1}{3}\sum_{\alpha}E^{2}_{\nu}A_{\nu_{\alpha}}(E_{\nu})=\frac{K_{\pi}}{4}E^{2}_{\gamma}A_{\gamma}(E_{\gamma})
\end{equation}
where E$_{\gamma}$ = 2E$_{\nu}$ and K$_{\pi}$ = 1 is a factor which accounts for the ratio of charged to neutral pions for $p\gamma$ interactions (for full derivation see e.g. \cite{Ahlers2018}). 

In the case of TXS-like sources we adapt the phenomenological model of \cite{Halzen2019}. The emerging gamma-ray flux is given by:
\begin{equation}
    \frac{dN_{\gamma}}{dE} = A_{\nu} E^{-2}e^{(-E_{L}/E -E/E_{H})}
\end{equation}
where E$_{L}$ and E$_{H}$ are low- and high-energy cutoffs, and A$_{\nu}$ is proportional to the simulated neutrino flux normalization. In a case of TXS 0506+056 located at redshift $z$ = 0.335, the E$_{L}$ = 0.1 TeV and E$_{H}$ = 20 TeV (see \cite{Halzen2019}). For the sources located at different redshifts, we scale those values accordingly.


\section{CTA simulations}
To simulate the CTA follow-up observations of the neutrino alerts we use the \texttt{ctools} package with \texttt{gammalib} \cite{ctools}. We employed the \texttt{prod3b-v2} CTA instrument response functions (IRFs) for the standard IRF set for the preliminary Omega Configuration, and the \texttt{prod3b} CTA IRFs for the preliminary Alpha Configuration sub-array and high night sky background (NSB) conditions. When possible, 1,000 FIRESONG alerts were considered for each of our density-luminosity sample using the different IRFs configurations. We obtain the redshift, spectrum normalization and declination for the alerts from running FIRESONG, while the right ascension is assigned randomly. For all sources, we take into account extragalactic background light (EBL) absorption \cite{Dominguez2011}.

The first set of IRFs considered corresponds to the preliminary North and South Omega configuration: 4 LSTs and 15 MSTs for CTA-N and 4 LSTs, 25 MSTs and 70 SSTs for CTA-S  \citep{Acharyya:2019}. The Omega IRFs set contains three zenith angle observation options at 20$^{\circ}$, 40$^{\circ}$ and 60$^{\circ}$; and it also accounts for the azimuth dependence coming from the magnetic field pointing direction: North, South or an average over the azimuth direction. The second IRFs set is the preliminary Alpha Configuration array, which was expected during the construction phase of CTA. A sub-array with a lower number of operational telescopes: 15 MSTs and 50 SSTs for CTA-S and 4 LSTs and 5 MSTs for CTA-N. Note that it differs from the current official Alpha Configuration which includes 4 LSTs + 9 MSTs for CTA-N and 14 MSTs + 37 SSTs for CTA-S.

For each alert in the Omega Configuration, we simulate the photon events list for 30 min of either North or South site observations with a 5.0$^{\circ}$ ROI centered at a source (with the tool \texttt{ctobsssim}). We consider the following energy ranges: 0.03-200 TeV for 20$^{\circ}$, 0.04-200 TeV for 40$^{\circ}$ and 0.110-200 TeV for 60$^{\circ}$ zenith angles (the lower and upper limits are defined by the IRFs). The energy dispersion effect was also considered in the simulated observations. We then perform a maximum likelihood fitting using the tool \texttt{ctlike} in an unbinned mode. The test statistic (TS) equal or higher than 25 qualifies as a source detection at the $\sim$5$\sigma$ level. For the Alpha configuration, zenith angle and azimuth pointing direction combinations were the same as in the Omega Configuration. \texttt{ctools} simulations for SFR and flat evolution steady sources followed the same specifications (ROI, observation times, zenith angles).

The average Light of the Night Sky (LoNS) on La Palma site is around $1.7 \times 10^{12}$ ph m$^{-2}$ sr$^{-1}$ s$^{-1}$ on the IACTs sensitive range \citep{MIRZOYAN:1998}. Measurements on both sites (CTA-N and CTA-S) showed similar NSB levels \citep{HASSAN:2017}. CTA observations up to 5 times the NSB level found in dark sky patches away from the Galactic Plane are anticipated when the moon is above the horizon. Simulations with our last set of IRFs tested this scenario, only 4 selected cases were at our disposal within the \texttt{prod3b} IRFs version: CTA-N 20$^\circ$ for North, South and average pointing direction in azimuth, and CTA-S 20$^\circ$ average in azimuth. 

The \texttt{ctools} simulations for the \textit{TXS-like} flaring sources followed the same scheme as in the steady source scenario. The previously described 3 IRFs sets were used. The IRF specifications such as energy range, ROI, zenith angles and magnetic field azimuth dependence were the same as in the steady source scenario.

\begin{figure}
    \centering
    \includegraphics[width=0.45\textwidth]{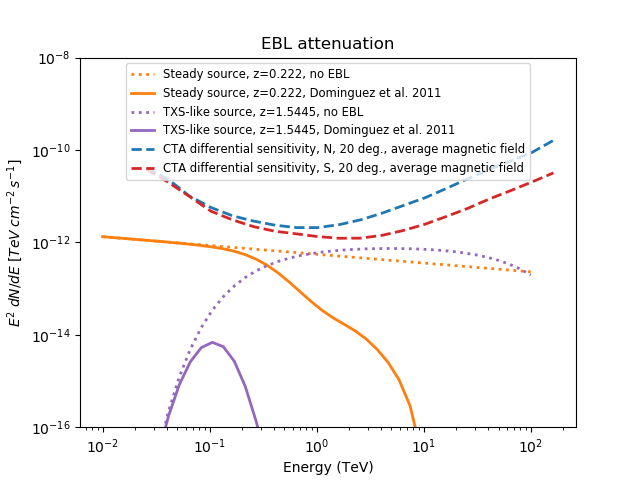}\includegraphics[width=0.45\textwidth]{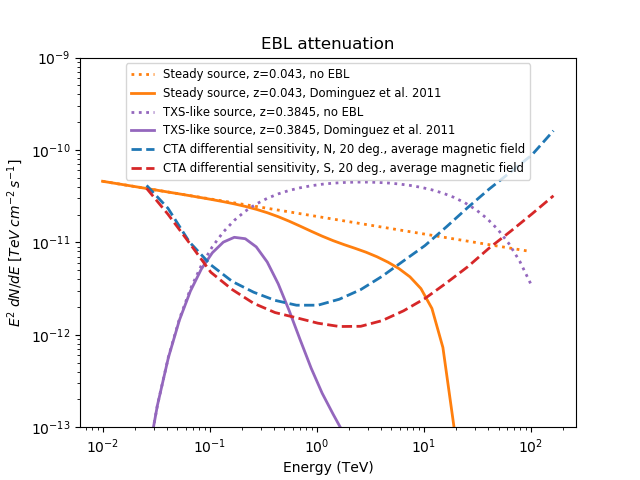}
    \caption{The energy spectra without and with the EBL attenuation vs the CTA differential sensitivity for the undetected (left) and detected (right) sources.}
    \label{fig:ebl}
\end{figure}

\section{Results}
\begin{figure}
    \centerline{
    \includegraphics[width=0.5\textwidth]{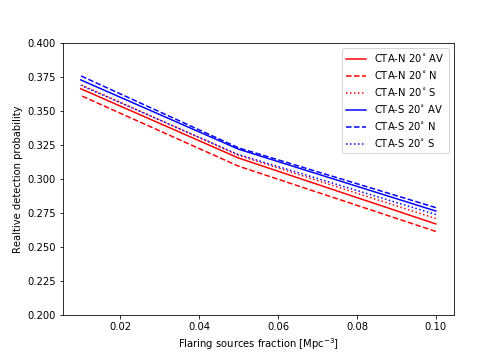}
    \includegraphics[width=0.5\textwidth]{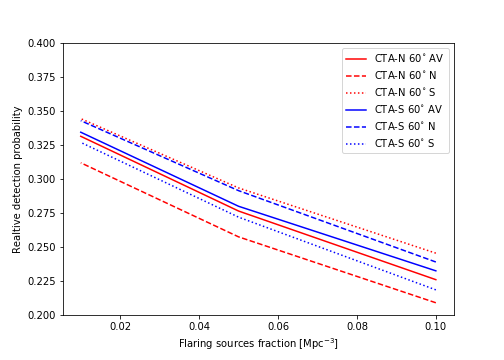}}
    \caption{Detection probability as a function of flaring sources fraction $F$ for 30 min observation time, assuming CTA-N (red) and CTA-S (blue) observations with zenith angle of 20$^{\circ}$, 
    and 60$^{\circ}$. The solid, dashed and dotted lines represent 3 azimuth angle alignments: average, North and South.}
    \label{fig:txs_like}
\end{figure}

\textbf{TXS-like flaring blazars.} Figure \ref{fig:txs_like} shows the detection probability as a function of flaring sources fraction $F$ for 30 min observation time for CTA-N and CTA-S, assuming different zenith and azimuth angle configurations. This probability, taking into account a background event contamination of 50\%, is highest for the population with 1\% of flaring sources. 

In this scenario, the detection probability is almost identical for CTA-N and CTA-S (difference $<$1\%) with average azimuth direction. There's also almost no difference for 20$^{\circ}$ and 40$^{\circ}$ zenith IRFs (again $<$1\%), but with higher zenith angles (60$^{\circ}$) we measure a decrease of $\sim$4\%. The influence of magnetic field is minimal, although slightly more evident for CTA-N and for high zenith angles.

\begin{figure}
    \centering
    \includegraphics[width=0.8\textwidth]{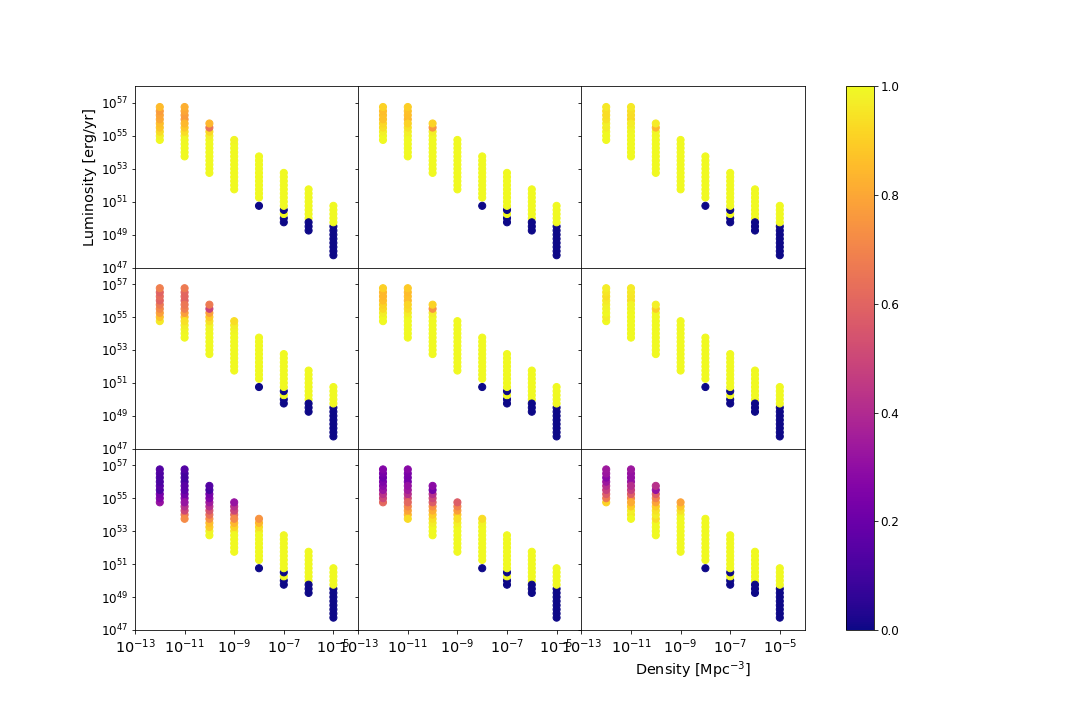}
    \caption{Detection probability (color scale) as a function of source luminosity ($L$) and local density ($\rho$) for sources following the SFR redshift evolution for 30 min observations with CTA-N. The plots in rows from top to bottom represent results for zenith angles of 20$^{\circ}$, 40$^{\circ}$ and 60$^{\circ}$, while the columns separate 3 azimuth angle alignments: North, average and South (from left to right). An abrupt drop in the detection probability from 1 to 0 occurs for $L$ and $\rho$ for which the FIRESONG output contained no alerts.}
    \label{fig:results_noevo_ctan}
    \end{figure}

 \textbf{Steady sources}. For the steady sources case we present the detection probability as a function of source luminosity ($L$) and local density ($\rho$).
 Fig. \ref{fig:results_noevo_ctan} shows our results for sources following the SFR evolution for 30 min observations with CTA-N and CTA-S respectively. The plots in rows from top to bottom represent results for zenith angle of 20$^{\circ}$, 40$^{\circ}$ and 60$^{\circ}$, while the columns separate three different azimuth angle alignments: North, average and South (from left to right).
 
With low to mid zenith angle observations (20$^{\circ}$-40$^{\circ}$) CTA-N will be able to detect all sources down to the density of $\rho$=10$^{-9}$ Mpc$^{-3}$. A drastic performance loss, up to 65\%, is measured at high zeniths (60$^{\circ}$). The effect of the geomagnetic field is reflected in the 10-30\% difference in detection probability for North/South azimuth directions for low to high zeniths. The CTA-S array shows a similar response as CTA-N (within 10\% for average azimuth). The main differences are a higher performance loss, up to 70\%, at high zeniths (60$^{\circ}$) and a smaller influence of geomagnetic field (5-15\% difference in detection probability for North/South azimuth directions for low to high zeniths). 

The higher NSB levels do not influence the detection probability for sources with $\rho>$10$^{-10}$ Mpc$^{-3}$ and $L<$10$^{55}$ erg/yr. For lower densities and higher luminosities the detection rate decreases by 10-15\%. This effect is similar for both considered arrays and cosmological evolution scenarios.

In the Alpha configuration used in this study the CTA-N performance is almost the same as for the full Omega array. On the other hand, the lack of LSTs in the Alpha configuration of the CTA-S drastically diminishes its detection capability. The effect is most visible for $\rho<$10$^{-9}$ and $L>$10$^{55}$ erg/yr. In the most extreme cases the detection probability drops by one third for zenith angles of 20$^{\circ}$, for 40$^{\circ}$ by half, and with 60$^{\circ}$ at most 10\% of sources are detected.


\begin{figure}
    \centering
    \includegraphics[width=0.45\textwidth]{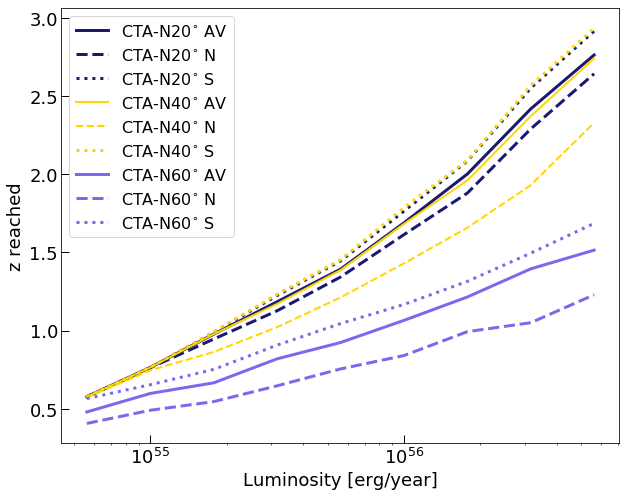}\includegraphics[width=0.45\textwidth]{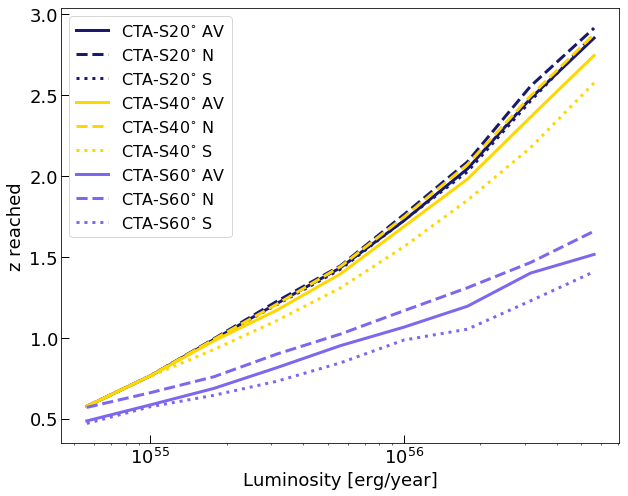}
    \caption{Redshift reach from CTA-N (left) and CTA-S (right) simulations for steady sources at density $\rho=10^{-12}$ Mpc$^{-3}$ following SFR evolution. Zenith angles of 20$^{\circ}$, 40$^{\circ}$ and 60$^{\circ}$ are considered. The solid lines represent the IRF with an average magnetic field, while the dashed (dotted) lines are for the pointing North (South) direction.}
    \label{fig:z-reached}
\end{figure}

We define the CTA redshift reach as the maximum redshift for 90\% of the detected sources in the distribution for each local density (the last decile was cut due to the low number of detected sources). Figure \ref{fig:z-reached} shows an example plot for the steady SFR evolution case at $\rho=10^{-12}$Mpc$^{-3}$ observed with CTA-N and CTA-S. The highest redshift reach among our simulations is given by this density and is up to $z\sim2.8$. For observations at 20$^{\circ}$ and 40$^{\circ}$ zenith angle the redshift reach is similar, however there is a big drop for 60$^{\circ}$. This is a general behaviour for all the density range considered. For sources with the flat redshift evolution the trends are similar as those described above, but the redshift reach is lower than for the SFR evolution.

\small
\section*{Acknowledgments}
KS, AB, ARdL and OS gratefully acknowledge financial support from the agencies and organizations listed here:                www.cta-observatory.org/consortium\_acknowledgments.
This work was performed within the CTA Transients Physics Working Group. KS acknowledge the the European Union’s Horizon 2020 Programme under the support of the European Union’s Horizon 2020 Programme under the AHEAD2020 project (grant agreement n. 871158). CFT and IT acknowledge support by NSF grant PHY-1505230. ARdL acknowledges the support of the National Council for Science and Technology from Mexico (CONACYT). 

This research made use of \texttt{ctools}, a community-developed analysis package for Imaging Air Cherenkov Telescope data. \texttt{ctools} is based on \texttt{GammaLib}, a community-developed toolbox for the scientific analysis of astronomical gamma-ray data.

\clearpage
\section*{Full Authors List: \Coll\ Collaboration}
%
%
\scriptsize
\noindent
%

\textbf{The Cherenkov Telescope Array Consortium July 2021 Authors}

H.~Abdalla\textsuperscript{1}, H.~Abe\textsuperscript{2},
S.~Abe\textsuperscript{2}, A.~Abusleme\textsuperscript{3},
F.~Acero\textsuperscript{4}, A.~Acharyya\textsuperscript{5}, V.~Acín
Portella\textsuperscript{6}, K.~Ackley\textsuperscript{7},
R.~Adam\textsuperscript{8}, C.~Adams\textsuperscript{9}
% ...

1 : Centre for Space Research, North-West University, Potchefstroom, 2520, South Africa

2 : Institute for Cosmic Ray Research, University of Tokyo, 5-1-5, Kashiwa-no-ha, Kashiwa, Chiba 277-8582, Japan

3 : Pontificia Universidad Católica de Chile, Av. Libertador Bernardo O'Higgins 340, Santiago, Chile

4 : AIM, CEA, CNRS, Université Paris-Saclay, Université Paris Diderot, Sorbonne Paris Cité, CEA Paris-Saclay, IRFU/DAp, Bat 709, Orme des Merisiers, 91191 Gif-sur-Yvette, France

5 : Centre for Advanced Instrumentation, Dept. of Physics, Durham University, South Road, Durham DH1 3LE, United Kingdom

6 : Port d'Informació Científica, Edifici D, Carrer de l'Albareda, 08193 Bellaterrra (Cerdanyola del Vallès), Spain

7 : School of Physics and Astronomy, Monash University, Melbourne, Victoria 3800, Australia

8 : Laboratoire Leprince-Ringuet, École Polytechnique (UMR 7638, CNRS/IN2P3, Institut Polytechnique de Paris), 91128 Palaiseau, France

9 : Department of Physics, Columbia University, 538 West 120th Street, New York, NY 10027, USA

% ...

\end{document}